# From Black Box to Glass Box: Cross-Model ASR Disagreement to Prioritize Review in Ambient AI Scribe Documentation

*Abdolamir Karbalaie[1], Fernando Seoane [1,3,4,5], Farhad Abtahi[1,2,3]*

[1]Department of Clinical Science, Intervention and Technology, Karolinska Institutet, 17177 Stockholm, Sweden
[2]Department of Biomedical Engineering and Health System, School of Engineering Sciences in Chemistry, Biotechnology and Health, KTH Royal Institute of Technology, 14157 Huddinge, Sweden
[3]Department of Clinical Physiology, Karolinska University Hospital, 17176 Stockholm, Sweden
[4]Department of Textile Technology, Faculty of Textiles, Engineering and Business Swedish School of Textiles, University of Borås, 503 32 Borås, Sweden
[5]Department of Medical Technologies, Karolinska University Hospital, 141 57 Huddinge, Sweden

## Abstract

**Objectives:** To evaluate whether cross-model disagreement among heterogeneous automatic speech recognition (ASR) systems can serve as a reference-free uncertainty signal for prioritizing human verification in medical transcription workflows. This is a methodological validation study; clinical accuracy was not evaluated.

**Materials** and Methods: We assembled a corpus of 50 publicly available medical education audio clips (total duration: 8 h 14 min) and ran each clip through eight ASR systems spanning commercial APIs and open-source engines. We constructed consensus pseudo-references from aligned multi-model outputs, characterized token-level agreement using a majority-strength metric, and analyzed disagreement composition by error type (content vs. punctuation/formatting). Per-model agreement and winner analyses were computed using leave-one-model-out (jackknife) consensus references. We assessed inter-model reliability using intraclass correlation coefficients (ICC) and examined variation across accent groups.

**Results:** Inter-model reliability was low (ICC[2,1] = 0.131), consistent with heterogeneous failure modes across systems. Among 76,398 token positions, 72.1% achieved low-risk agreement with the consensus pseudo-reference (i.e., 7–8 models concordant), while 2.5% fell into high-risk categories (i.e., 0–3 models concordant), this risk was unevenly distributed, ranging from 0.7% to 11.4% across accent groups. Low-agreement regions showed enrichment for content-category disagreements: the content fraction increased from 53.9% to 73.9% across quintiles of high-risk mass.

**Discussion:** Cross-model disagreement provides a sparse, localizable signal that is relatively more concentrated in lexical divergence within meaning-bearing categories than in superficial formatting differences, supporting its potential use as a triage mechanism for prioritizing review.

**Conclusions:** Multi-model disagreement offers a practical, human-annotation-free approach for surfacing transcription uncertainty in medical speech, enabling targeted verification without requiring human-verified reference transcripts.

**Keywords:** automatic speech recognition; clinical documentation; uncertainty quantification; ensemble methods; health informatics

**Lay Summary**

When clinicians use speech recognition software to draft clinical notes, errors can go unnoticed without careful review. This study tested whether comparing outputs from multiple speech recognition systems could help identify which parts of a transcript are most likely to need careful review. We found that when different systems disagree about what was said, those disagreements tend to involve meaning-bearing words (content) rather than minor formatting differences like punctuation. This approach could help healthcare workers focus their review efforts on the parts of a transcript most likely to warrant double-checking, though further studies are needed to confirm that flagged regions correspond to actual errors in clinical settings.

**Introduction**

Ambient artificial intelligence (AI) scribes that record patient encounters and generate visit notes are increasingly being deployed, motivated by their potential to reduce documentation burden and physician burnout [1][2] [3]. A recent randomized trial of 238 physicians across 72,000 patient encounters demonstrated that AI scribes can reduce documentation time by up to 9.5% and improve burnout scores, providing rigorous evidence of their operational benefits [4]. As healthcare systems adopt these tools at scale, the underlying automatic speech recognition (ASR) component becomes a key dependency for clinical documentation.

However, transcription reliability remains a persistent concern. Even as ambient AI scribes demonstrate efficiency gains, clinicians have reported that AI-generated notes occasionally contain clinically significant inaccuracies, including omissions and attribution-related failures [4][5]. A recent policy analysis emphasized that such tools require ongoing vigilance and active physician review rather than passive acceptance [1]. The operational challenge therefore extends beyond generating a transcript to prioritizing human verification, given that clinically oriented ASR systems still exhibit non-trivial error rates in patient–clinician conversations [6][7].

A major barrier to systematic quality assurance is that standard ASR evaluation assumes access to a human-verified reference transcript, typically to compute aggregate metrics such as word error rate (WER) [7]. In many practical settings especially at scale (e.g., continuous monitoring across thousands of encounters) human-verified reference transcripts are unavailable, limiting routine auditing [7][8]. This limitation is compounded by the fact that aggregate error metrics collapse heterogeneous phenomena, conflating meaning-bearing lexical substitutions with superficial differences such as punctuation or formatting, motivating error-type-specific and semantic evaluation in clinical ASR [9]. Although some ASR systems expose confidence-like signals, these are inconsistently available across vendors. Where confidence measures are provided, they may be poorly calibrated for actionable, token-level decision support, leaving end-users with limited guidance on where careful checking is warranted [10][11].

In this study, we test whether disagreement among independent ASR systems can serve as a reference-free (no human-verified reference transcript), token-level signal for prioritizing human review in medical transcription. Instead of estimating correctness relative to an external gold standard, a multi-model approach treats independent ASR systems (which can exhibit complementary failure modes) as raters of the same underlying speech signal: strong agreement suggests transcript stability, whereas divergence flags candidate spans for adjudication [12]. This concept aligns with classical ASR system-combination and lattice consensus frameworks, which operationalize agreement structure across competing hypotheses [12] [13]. Here, rather than combining systems solely to minimize average error, we use *disagreement* as an explicit triage signal to localize where manual verification is likely to yield the highest safety value. This approach aligns with the recently proposed MEDLEY (Medical Ensemble Diagnostic system with Leveraged diversitY) framework, which treats cross-model disagreement as a reference-free uncertainty signal for targeted verification [14].

For such a signal to be clinically useful, two requirements must hold. First, low-agreement regions should preferentially capture meaning-bearing disagreements rather than trivial surface-form variation; otherwise, such a tool could inflate review burden without commensurate value [9]. Second, because ASR performance is known to vary with disfluency, demographic characteristics, accent, and multilingual settings, any disagreement-based triage signal should be evaluated for robustness and equity across these sources of variability [15][16] [17].

The objective of this study was to assess whether multi-ASR agreement can function as a practical, annotation-free signal of transcription uncertainty for clinical review. The central question is whether disagreement among independent ASR systems is (1) sparse, (2) localizable, and (3) enriched for meaning-bearing lexical divergence, because these properties are required for a practical review triage signal. This is a methodological validation study using a reproducible medical-speech dataset; we do not evaluate clinical accuracy against gold-standard transcripts. We addressed four research questions: (1) Are agreement-based uncertainty estimates stable under different reference-construction choices? (2) Are low-agreement regions systematically enriched for content-category disagreements? (3) How do agreement and risk profiles vary across accent or demographic groups? (4) To what extent are disagreement patterns associated with recording quality versus linguistic variability?

## Materials and Methods

### Study Design and Data Collection

We compiled an English-language medical speech corpus from publicly available YouTube videos selected to reflect clinically relevant communication styles, including medical education narration and dictation-like speech. Each source video was segmented into approximately 10-minute audio clips, yielding N = 50 clips as the primary unit of analysis. For each clip, we retained the audio segment and associated non-identifying metadata (i.e., accent label, speaking rate, and acoustic properties). Speaking rate (words/min) and word counts were derived from the Gemini Flash 2.5 transcript by dividing the transcript word count by clip duration (minutes). This measure is used as a descriptive proxy and is not a human reference transcript. The dataset follows prior practice of constructing publicly available, web-sourced speech corpora for ASR development and benchmarking (e.g., GigaSpeech and MediaSpeech) [18][19].

Ethical requirements were assessed in accordance with institutional policy for secondary analyses of publicly available materials. No participant recruitment, intervention, or direct interaction was performed. To minimize privacy risks, analyses were reported at the clip and subgroup levels; speaker names and identifying fields were excluded from analytical tables; and no protected health information or personally identifiable information was collected or analyzed.

## Accent Annotation and Verification

Accent labels were obtained primarily from video metadata (e.g., speaker biography, affiliation, or stated region). To reduce labeling noise and harmonize categorization, we performed a secondary verification step using Google Gemini 2.5 Flash for metadata harmonization only, n
ot for primary analysis. All Gemini outputs were manually adjudicated for discordant or low-confidence cases, and final accent annotations were mapped to a standardized taxonomy through deterministic rules. Additional details are provided in Supplementary Method S1.

## Audio Extraction and Standardization

All recordings were originally distributed as video files with heterogeneous audio encoding. For each video, the audio stream was demultiplexed and decoded, and a fixed-length analysis segment targeting 600 seconds was extracted (or the full duration if the source was shorter). All processed clips were stored as mono WAV at 16 kHz sampling rate and 256 kbps bitrate, ensuring that every ASR system received acoustically identical input. Detailed technical characteristics of the original source video audio and the standardized analysis segments are provided in Supplementary Method S2. All ASR systems were run contemporaneously using identical audio files to ensure reproducibility.

## ASR Systems

Each clip was processed through eight distinct ASR systems to quantify transcription uncertainty via cross-model disagreement (**Table 1**). The ensemble spanned different architectural families and deployment contexts, chosen to increase diversity in failure modes and thereby render disagreement signals informative rather than redundant. We intentionally included a legacy pair (Wav2Vec 2.0 [20], NeMo QuartzNet[21]) to widen architectural diversity and to probe sensitivity to ensemble composition; however, we report reliability for a modern subset (Whisper Turbo v3, Whisper Large v3, Gemini Flash 2.5, Vox Mini, Speechmatics, MedASR) to reflect a more deployment-realistic ensemble. Exact model endpoints, version identifiers, and inference/processing settings are listed in Supplementary Table S3. All systems were run with default inference settings to reflect typical deployment without domain adaptation.

## Transcript Normalization and Tokenization

Before alignment, all transcripts underwent uniform normalization to prevent false disagreement from formatting differences. Transcripts were tokenized by splitting on whitespace, then normalized according to deterministic rules applied uniformly across all systems. Normalization included case conversion (lowercase), punctuation stripping, numeric normalization (mapping word forms to digits), and contraction standardization. A curated filler word vocabulary enabled separate classification of filler-only differences. These rules ensured that reported disagreements reflect genuine acoustic or semantic differences rather than formatting artifacts. The complete normalization rules and mapping resources are provided in Supplementary Methods S4 and Tables S4–S6.

**Aligned Transcript Representation and Pseudo-Reference Modes**

We aligned the eight normalized model transcripts for each clip into a shared token coordinate system, yielding an aligned token matrix in which rows correspond to ASR systems and columns correspond to aligned positions. Unless otherwise stated, analyses are reported over evaluated aligned positions, excluding all–gap columns (i.e., positions where all systems contain a gap). We evaluated outputs under three deterministic reference-construction modes: (i) Consensus (primary), which selects a consensus token at each aligned position by majority vote; (ii) Centroid, which selects a single existing model transcript that is most central among hypotheses; and (iii) Single-model reference, which treats a designated model output as the reference for sensitivity analysis. Full alignment parameters, tie-handling rules, and mode-specific implementation details are provided in Supplementary Methods S5.

**Jackknife (leave-one-model-out) consensus for per-model scoring.**

To mitigate circularity when scoring a model against an ensemble-derived pseudo-reference, we used a leave-one-model-out (jackknife) procedure for all per-model agreement and winner analyses [22]. For each clip and evaluated model m, we constructed a jackknife consensus pseudo-reference from the remaining K−1 systems (where K denotes the number of ASR systems) using the same Consensus voting rule (Supplementary Method S5). We then scored model m against its jackknife reference over non-gap reference positions using percent-identical and content-difference definitions.

**Error Taxonomy**

For each clip and evaluation instance, we compared each model hypothesis to the selected reference token-by-token using the aligned transcript matrix. Each evaluated position was assigned exactly one label using a deterministic decision order, yielding a mutually exclusive taxonomy: (1) Identical, (2) Punctuation, (3) Contraction, (4) Numeric, (5) Filler, and (6) Content (all remaining lexical mismatches, including meaning-bearing differences). Full labeling rules and decision order are provided in Supplementary Method S6. Throughout, 'content-category' disagreement denotes lexical category (not verified clinical correctness) and therefore supports enrichment analyses rather than accuracy claims.

**Derived Metrics and Risk-Band Definitions**

Using clip-level aligned comparisons to the selected reference (Consensus by default), we computed two primary metrics for each clip × model: percent-identical agreement ($p_{identical}$) and content-difference rate ($r_{content}$). Let $N_{total}$ denote the number of evaluated aligned positions for a clip (reference token non-gap; excluding all-gap columns). We define

$$p_{identical} = \frac{N_{ident}}{N_{total}}, \qquad r_{content} = \frac{N_{content}}{N_{total}},$$

where $N_{ident}$ counts positions where the model's normalized token exactly matches the reference and $N_{content}$ counts positions labeled as content differences under the disagreement taxonomy (excluding punctuation, contraction, numeric, and filler mismatches). Unless stated otherwise, corpus summaries are computed in two ways: (i) per-model, by averaging each model's clip-level metrics across clips; and (ii) as reference-mode 'mean agreement', by averaging percent-identical across models within each clip and then averaging across clips (reporting the standard deviation across clips).

To quantify ensemble support at each aligned position $i$, we computed majority strength $A_i \in \{0,..., 8\}$, defined as the number of ASR systems whose normalized token matches the reference token at $i$ (gaps treated as non-matches). For 8 systems, we pre-specified three risk bands: low-risk $A_i \in \{7, 8\}$, medium-risk $A_i \in \{4, 5, 6\}$, and high-risk $A_i \leq 3$.

## Statistical Analysis

Inter-model reliability was assessed by treating ASR systems as raters and clips as targets, computing single-measure intraclass correlation coefficients for absolute agreement (ICC[2,1]) separately for percent-identical and content-difference rate. In addition to the full ensemble, we report subgroup reliability for a modern subset (k = 6) and a legacy pair defined as Wav2Vec 2.0 and NeMo QuartzNet, to illustrate how agreement changes under different system compositions. Pairwise Spearman rank correlations were computed between per-clip score vectors for each model pair to identify systems with similar clip-wise behavior. We also computed mean paired differences in clip-level percent-identical between model pairs, defined as $\Delta p_{identical} = p_{identical}(model(m)) - p_{identical}(model(n))$ (percentage points), averaged over clips. Associations between clip-level audio covariates and disagreement signals were assessed using Spearman correlations, including correlations between clip duration, loudness, and estimated SNR proxies and the clip-level content-difference rate. Expanded reliability and correlation analysis details are provided in Supplementary S7.

## Results

### Corpus Characteristics

All model-wise summaries below quantify agreement with the consensus pseudo-reference or its jackknife variant to characterize inter-model heterogeneity relevant to disagreement-based triage. The evaluation corpus comprised 50 audio clips with a total duration of 29,635 seconds (approximately 8 hours and 14 minutes). Mean clip duration was 592.5 seconds (SD = 53.2; range: 223.9–612.0). Mean speaking rate was 158.1 words per minute (SD = 30.9). The corpus included speakers from 18 accent categories, with Singaporean (n = 10), American (n = 8), Australian (n = 8), and British (n = 6) being most represented. Gender distribution was 56% male and 44% female. Descriptive statistics of the evaluation corpus and ASR ensemble are summarized in **Table 1**.

Table 1. Descriptive statistics of the evaluation corpus and ASR ensemble (N = 50 clips)

| Category | Measure | Value |
|---|---|---|
| **Audio Corpus** | Number of clips | 50 |
| | Total duration | 29,635.10 s (≈ 8 h 14 min) |
| | Clip duration(mean ± SD) | 592.5 ± 53.2 s |
| | Clip duration (range) | 223.9 – 612.0 s |
| | Sample rate | 16 kHz (all clips) |
| | Channels | Mono (1 channel; all clips) |
| | Bitrate | 256 kbps (all clips) |
| | Loudness (mean ± SD) | −25.18 ± 5.43 dBFS (RMS proxy) |
| **Speech Content** | Words per clip (mean ± SD) | 1,560 ± 340 |
| | Words per clip (range) | 579–2,292 |
| | Speaking rate (mean ± SD) | 158.1 ± 30.9 words/min |
| | Speaking rate (range) | 92–229 words/min |
| **Speakers** | Videos/speakers | 50 (one speaker per clip) |
| | Gender | 28 male (56%), 22 female (44%) |
| | Accent distribution (examples) | 10 Singaporean, 8 American, 8 Australian, 6 British, 3 Indian, 2 Arabic, 2 New Zealand, plus 1 each of French, German, Japanese, Korean, Nigerian, Spanish, Norwegian, Swedish, South African, Canadian, Filipino |
| **ASR Ensemble** | Systems (n) | 8 (Gemini Flash 2.5, Vox Mini, Speechmatics, Whisper Turbo v3, Whisper Large v3, MedASR, Wav2Vec 2.0, NeMo QuartzNet) |
| | Total aligned tokens (across all clips and models) | 696,204 |
| | Evaluated aligned positions (Consensus mode) | 76,398 |

**Note**: Word counts and WPM are derived from the Gemini Flash 2.5 transcript word count divided by the measured clip duration (not from a human reference transcript). Accent labels were derived from public metadata and standardized to a cleaned taxonomy after automated verification and manual adjudication. "Total aligned tokens" denotes the sum of normalized token counts across all 8 ASR outputs and all clips. "Evaluated aligned positions (Consensus mode)" denotes the number of Consensus pseudo-reference token positions used as the denominator for agreement/risk-band analyses (excluding all–gap columns and positions where the reference token is a gap).

Table 2. Per-Model Agreement With Jackknife (Leave-One-Model-Out) Consensus Pseudo-Reference.

| ASR System | Percent-identical mean (SD) | Content-difference % mean (SD) |
|---|---|---|
| Vox Mini | 93.3 (3.2) | 3.5 (2.5) |
| Gemini Flash 2.5 | 92.7 (3.1) | 2.7 (2.3) |
| Whisper Turbo v3 | 92.5 (4.0) | 3.6 (3.1) |
| Speechmatics | 91.7 (2.9) | 2.8 (1.8) |
| Whisper Large v3 | 91.5 (4.2) | 4.7 (3.7) |
| Wav2Vec 2.0 | 77.4 (7.0) | 13.5 (7.6) |
| MedASR | 72.7 (12.2) | 22.3 (12.6) |
| NeMo QuartzNet | 72.4 (9.2) | 19.0 (10.2) |

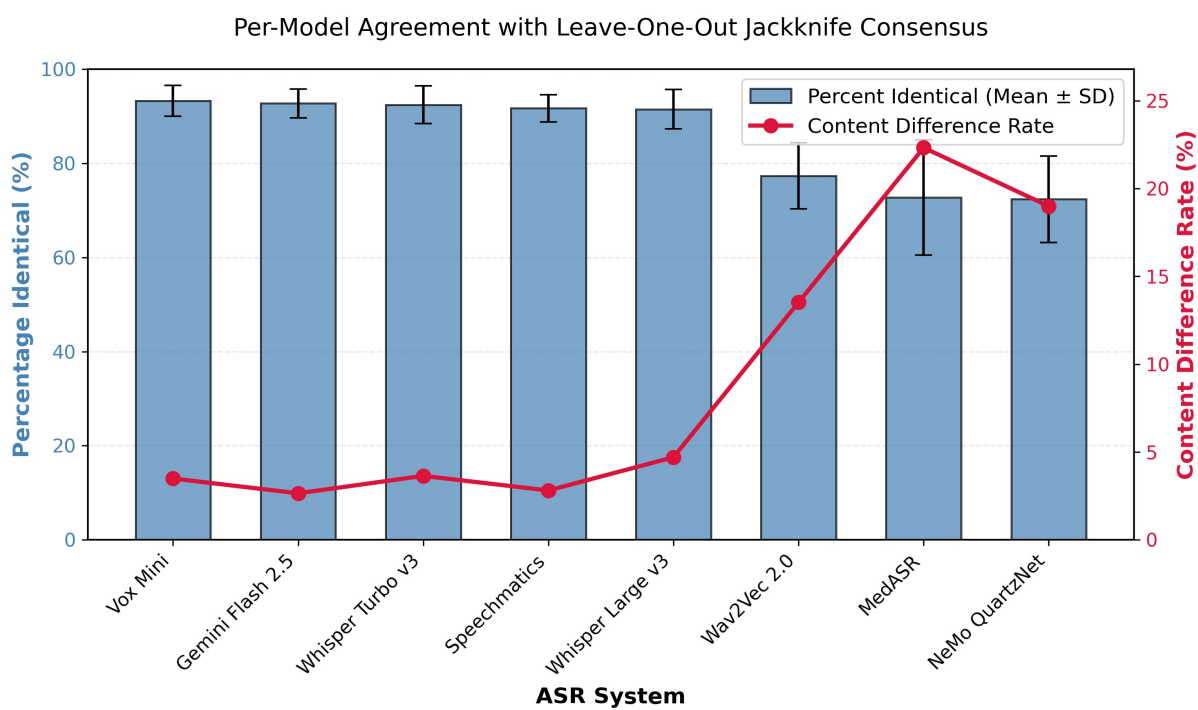

**Figure 1. Per-model agreement under the jackknife consensus pseudo-reference.** Mean percent-identical (±1 SD) across N = 50 clips, with an overlay showing mean content-difference rate per model. Higher bars indicate greater agreement with the ensemble-derived reference; lower content-difference rates indicate fewer meaning-bearing lexical divergences.

## Per-Model Agreement with the Jackknife Consensus Pseudo-Reference

Mean percent-identical under jackknife (leave-one-model-out) consensus differed markedly across models (Table 2) and is visualized alongside content-difference in Figure 1. Vox Mini achieved the highest mean agreement (93.3%; SD = 3.2), whereas MedASR and NeMo QuartzNet showed substantially lower agreement (72.7%; SD = 12.2; and 72.4%; SD = 9.2, respectively). Mean content-difference rates followed a corresponding gradient, ranging from 2.7% (Gemini Flash 2.5) to 22.3% (MedASR).

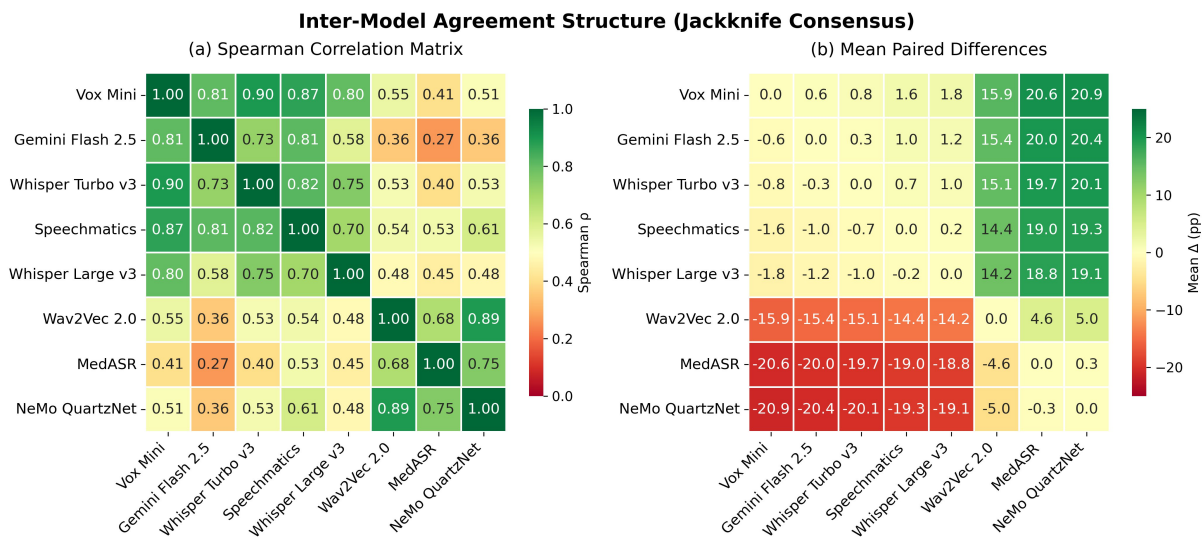

**Figure 2. Inter-model agreement structure and systematic offsets in clip-level percent-identical performance (jackknife consensus scoring; leave-one-model-out).** Panel (a) shows the pairwise Spearman rank correlation ($\rho$) matrix of per-clip percent-identical scores across the evaluated ASR models, computed over $N = 50$ clips under under jackknife consensus scoring. Panel (b) shows the corresponding mean paired differences $\Delta$ in $p_{identical}$ (percentage points), where each cell reports row − column averaged across the same clips; positive values indicate higher average $p_{identical}$ for the row model. Diagonal entries are 1.00 in (a) and 0 in (b).

## Winner Analysis and Inter-Model Reliability

Under jackknife evaluation, wins were distributed across four systems: Vox Mini (27/50; 54%), Gemini Flash 2.5 (13/50; 26%), Whisper Turbo v3 (7/50; 14%), and Whisper Large v3 (3/50; 6%). Winner counts stratified by accent group are shown in Supplementary Figure S1.

Across clips, jackknife percent-identical score vectors showed moderate between-model correlation (pairwise Spearman range 0.27–0.90, median 0.56). Inter-model similarity structure and systematic percent-identical offsets are summarized in Figure 2. Absolute-agreement reliability remained low for the full ensemble (ICC[2,1] = 0.131 for percent-identical; ICC[2,1] = 0.155 for content-difference rate). Reliability was lower within the modern subset (k = 6; ICC[2,1] = 0.114 for percent-identical; 0.088 for content-difference rate) but high within the legacy pair (k = 2; ICC[2,1] = 0.715 and 0.703, respectively).

## Majority-Strength Distribution

Given the observed variability in system behavior, we next examined whether disagreement is widespread or concentrated in a small subset of tokens. Among 76,398 evaluated aligned positions, the majority-strength distribution was concentrated at high agreement levels: 55.7% of positions achieved full agreement ($k = 8$), whereas positions with very low agreement were rare ($k \leq 2$: 0.5%). Using the pre-defined risk-band mapping (high-risk: $A_i \leq 3$), the corpus-level composition was 72.1% low-risk, 25.4% medium-risk, and 2.5% high-risk. The overall majority-strength histogram is shown in **Figure 3**.

Accent-stratified majority-strength distributions are shown in Figure 3, and accent-wise risk-band percentages with denominators ($n_{clips}$, $n_{tokens}$) are reported in Supplementary **Table**

S7. Among higher-representation groups, low-risk proportions were 76.6% (American; $n = 8$), 75.9% (Australian; $n = 8$), and 69.2% (Singaporean; $n = 10$).

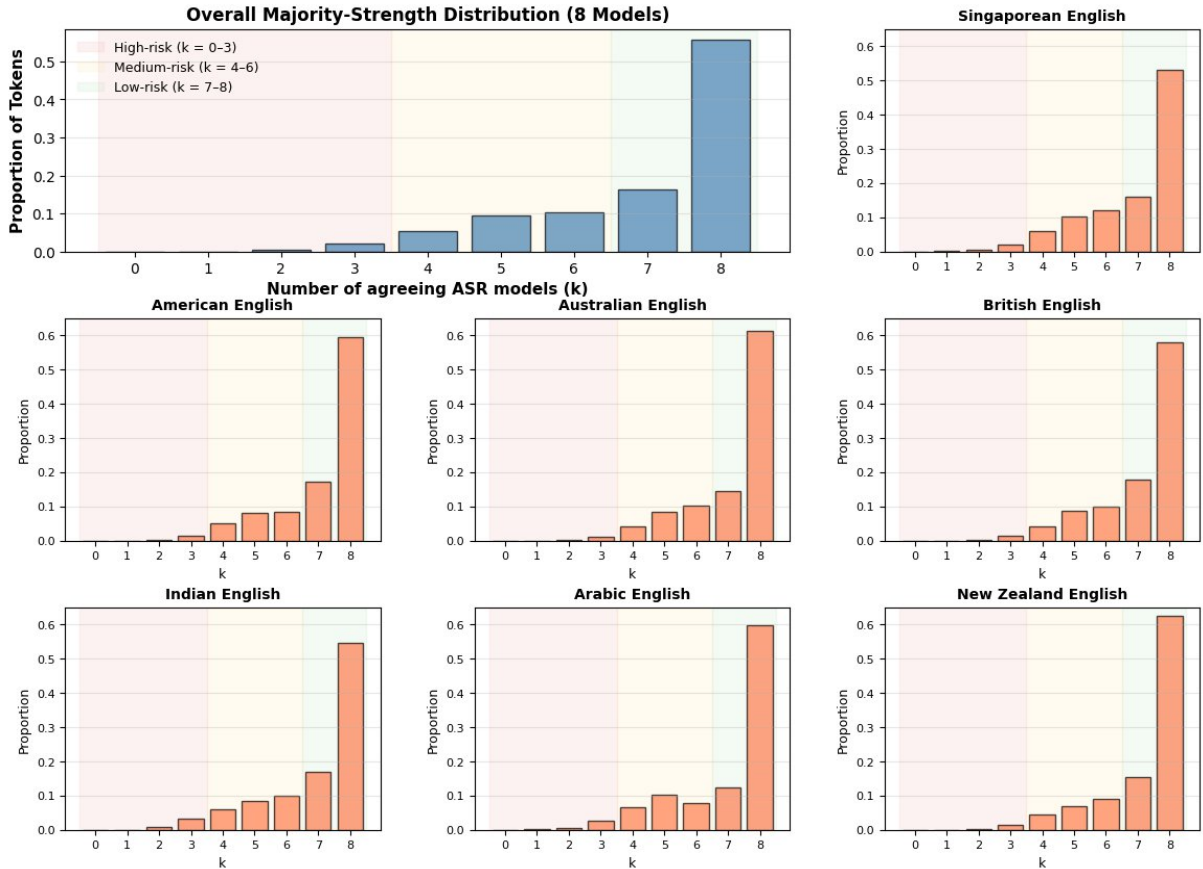

**Figure 3. Majority-strength distributions overall and by accent group (K = 8; Consensus mode).** Histograms show the distribution of agreement count $A_i \in \{0,...,8\}$. Risk bands correspond to high ($A_i \in \{0, 1, 2, 3\}$), medium ($A_i \in \{4, 5, 6\}$), and low ($A_i \in \{7, 8\}$) uncertainty levels. Accent-stratified results are exploratory given small group sizes.

## Content Enrichment in Low-Agreement Regions

Disagreement composition differed across quintiles of high-risk mass ($p_{high}$; **Table 3**). The mean content fraction increased from 53.9% in the lowest quintile to 73.9% in the highest quintile, whereas the mean punctuation fraction decreased from 45.3% to 25.3% (**Figure 4**). At the clip level ($N = 50$ clips), $p_{high}$ was positively correlated with content fraction (Spearman $\rho = 0.559$, $P < 0.001$; 95% CI [0.333, 0.730]) and negatively correlated with punctuation fraction ($\rho = -0.547$, $P < 0.001$; 95% CI [-0.725, -0.315]), the bootstrap CIs reflect moderate uncertainty at this sample size, and subgroup inference is therefore limited. In absolute terms, $p_{high}$ remained small even in the highest-risk quintile (mean 5.9%; maximum 11.4%).

## Reference Construction Mode Comparison

We compared the proposed Leave-One-model-Out (Jackknife) consensus against three baseline reference methods: non-jackknife consensus, Centroid and Single-model references. Across clips, mean percent-identical agreement was 85.5% (SD 4.6%) under the jackknife consensus, 87.2% (SD 4.2%) under the non-jackknife consensus, 86.6% (SD 4.6%) under the Centroid reference, and 81.3% (SD 5.6%) under the Single-model reference.

Table 3. Disagreement Composition by Quintiles of High-Risk Mass

| Quintile | N clips | High-risk mass, % (range) | Content, % | Punctuation, % |
|---|---|---|---|---|
| 1 (Lowest) | 10 | 0.7 (0.3-1.1) | 53.9 | 45.3 |
| 2 | 10 | 1.5 (1.3-1.6) | 57.6 | 41.3 |
| 3 | 10 | 2.1 (1.8-2.4) | 63.3 | 35.5 |
| 4 | 10 | 3.1 (2.5-3.5) | 63.4 | 35.8 |
| 5 (Highest) | 10 | 5.9 (3.6-11.4) | 73.9 | 25.3 |

**Note:** Quintiles were formed by ranking clips by high-risk mass $p_{high}$ (Consensus mode; high-risk defined as ajority strength $A_i \leq 3$) and splitting into five equal-sized groups ($n = 10$ clips per quintile). "High-risk mass" is the percentage of evaluated aligned positions in the clip with $A_i \leq 3$. "Content fraction" and "punctuation fraction" are computed within the set of labeled differences for each clip (fractions may sum to <100% when other categories such as Contraction/Numeric/Filler are present) and are reported as quintile means.

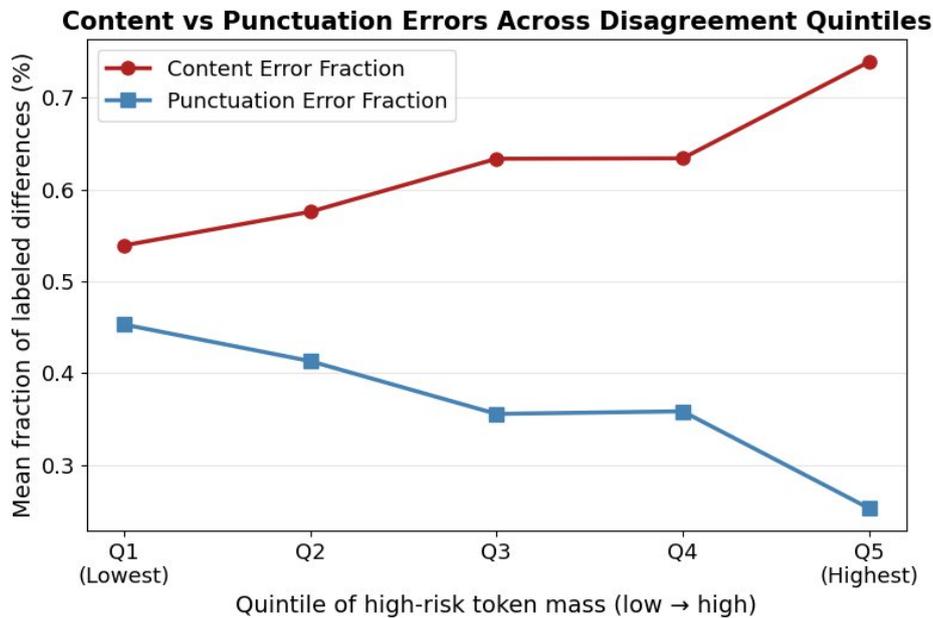

Figure 4. Error composition across quintiles of high-risk token mass (Consensus mode; high-risk defined as $\mathbf{A_i} \leq 3$). Points show quintile means of the fraction of labeled differences attributed to Content and Punctuation categories. Quintiles are defined by high-risk token mass (percentage of evaluated aligned positions with $A_i \leq 3$).

**Audio Quality Sensitivity**

Finally, we examined whether disagreement patterns were associated with clip-level audio quality variation. we computed a clip-level mean content-difference rate by averaging per-model content-difference rates across the ensemble and correlated this quantity with audio covariates; associations were weak (|ρ| ≤ 0.22). The largest magnitude association was observed for speaking rate (ρ = −0.220; P = 0.125). Clip duration, loudness, and estimated signal-to-noise ratio showed near-zero associations (|ρ| < 0.07).

## Discussion

This methodological study shows that cross-model disagreement across heterogeneous ASR systems can serve as a practical, no-reference signal of transcription uncertainty for medical speech review. Across the study's research questions, four observations are most consequential.

First, clip-level agreement was imperfect and model-dependent, low inter-model agreement suggests complementary failure modes rather than redundant outputs from the same recognizer family.

Second, disagreement was sparse and localizable: the majority-strength distribution concentrated at high agreement levels, with a small high-risk tail (defined as positions where ≤3 of 8 models match the reference token), supporting the feasibility of directing attention to a limited set of spans rather than requiring full-transcript review. Importantly, the absolute prevalence of risk bands is likely corpus-dependent. The YouTube medical-education clips used here are comparatively "clean" ASR inputs (single-speaker, prepared speech, limited background noise), whereas real patient–clinician encounters often include turn-taking, interruptions, overlapping speech, disfluencies, far-field capture, and occasional code-switching, all of which can increase recognition difficulty and divergence across systems [7]. Under those conditions, the disagreement tail may thicken, implying that the 2.5% high-risk rate observed here could underestimate the review burden in deployment. Practically, this means disagreement thresholds should be calibrated to the target setting (e.g., visit type, microphone setup, specialty, and patient population) rather than transferred verbatim from this corpus.

Third, when disagreement occurred, it was not uniformly "surface noise": low-agreement regions showed a systematic shift in composition toward content-category differences (meaning-bearing lexical divergence) rather than punctuation-only variation.

Fourth, exploratory accent-stratified summaries suggested that disagreement burden may vary across accent groups, but these estimates were unstable given small and uneven group sizes; associations with broad audio covariates were weak, suggesting disagreement is not primarily explained by clip-level quality differences within this corpus.

### Clinical Implications and Methodological Validation

These findings are timely given the rapid deployment of ambient AI scribes in clinical practice [1]. A recent randomized trial demonstrated that such tools can reduce documentation time and improve burnout measures, but also revealed that AI-generated notes occasionally contain clinically significant inaccuracies[4]. These findings underscore the need for active physician oversight [1][4]. Similarly, a policy analysis warned of the risks of passive acceptance of AI-generated documentation and called for governance guardrails including disabling auto-accept features and requiring active review of diagnoses and billing elements [1]. The multi-model disagreement signal evaluated here directly addresses this need: by surfacing transcript regions where ASR systems diverge, it provides a system-agnostic basis for prioritizing review without requiring reference transcripts or modifying individual ASR systems.

The low inter-model reliability (ICC[2,1] ≈ 0.13–0.16) observed in the results reflects non-interchangeable system behavior, suggesting that different modern architectures (e.g., Gemini vs. Whisper) do not consistently struggle on the same segments. Conversely, the high reliability within the legacy pair is consistent with more similar failure modes within that pair. The moderate pairwise correlations indicate that while models share some sensitivity to clip-level difficulty, their specific error patterns remain distinct. This heterogeneity supports the premise that systems contribute complementary failure modes rather than redundant rankings, making disagreement an informative signal. Ensemble composition is therefore a design choice with two competing goals:

(i) maximizing diversity to surface ambiguity (which can increase disagreement), and

(ii) matching the set of systems that would realistically be deployed.

While including weaker or legacy systems can increase the quantity of disagreement, our modern-subset results still show low reliability (k=6 ICCs), indicating that disagreement persists even among stronger systems and remains a viable triage signal. That said, the risk-band prevalence and optimal thresholds are expected to change with the chosen ensemble and should be recalibrated for the intended deployment configuration. Consensus and Centroid reference construction produced near-identical summaries, whereas Single-model reference construction reduced agreement and increased content divergence, supporting the robustness of ensemble-based pseudo-reference construction for monitoring workflows, with additional confirmation from leave-one-model-out (jackknife) scoring to reduce circularity in per-model evaluation.

MedASR showed the lowest agreement and highest content-difference rate (Table 2), despite being domain-labeled. This is not a clinical accuracy claim because the comparator is an ensemble pseudo-reference, not expert ground truth. The pattern may reflect domain/style mismatch with narrated medical education audio or default decoding choices that produce different lexical renderings. Disambiguating these explanations requires expert-adjudicated references and controlled inference settings.

A disagreement-based triage signal is practically useful only if it both enriches for meaning-bearing content and keeps review workload manageable. When clips were stratified by high-risk mass, the fraction of labeled differences attributed to content increased while punctuation decreased, indicating that low-agreement regions disproportionately capture lexical divergence in meaning-bearing word classes rather than superficial formatting changes. Even in the highest-risk quintile, high-risk tokens accounted for only a small share of positions (mean 5.9%), supporting the premise that a disagreement-based interface can direct attention to a limited subset of spans instead of requiring full-transcript review. In this workflow, the agreement threshold functions as a clinician-controlled review knob: inspecting only high-risk positions (A≤3) restricts attention to a very small proportion of tokens, whereas expanding to high+medium positions (A≤6) trades additional coverage for a still substantial reduction in review burden compared with full transcripts. A threshold sweep across A≤3–6 is summarized in Supplementary Table S8.

**Distinguishing Lexical Disagreement from Clinical Error**

A critical distinction must be drawn between content-category disagreement (i.e., lexical divergence in meaning-bearing word classes) and verified clinical error. Our taxonomy classifies disagreements by lexical category, not by clinical correctness. Accordingly, this study does not estimate the positive predictive value of a given risk band (e.g., the probability that an A≤3 token is clinically incorrect), because we did not perform expert adjudication against a human reference. The present evidence is therefore limited to enrichment (what kinds of differences concentrate in low-agreement regions), not correctness. A token flagged as content disagreement may reflect (i) a true transcription error by one or more systems, (ii) acceptable variation among plausible renderings, or (iii) downstream normalization effects. Conversely, clinically important errors can also occur under high agreement (e.g., if all systems converge on the same incorrect term), which would not be flagged by disagreement. Establishing whether disagreement enrichment translates into higher rates of clinically meaningful errors requires prospective evaluation with expert-adjudicated references.

Exploratory analyses suggested that disagreement burden varies across accent groups, though these findings must be interpreted cautiously given small and uneven sample sizes. A disagreement-based workflow could potentially make robustness gaps visible as differences in verification burden, offering a candidate approach for screening disparities in contexts where reference transcripts are unavailable, though validation against expert review remains necessary [15][16][23]. The weak associations between audio quality covariates and disagreement metrics suggest that disagreement patterns reflect linguistic or phonetic difficulty more than gross signal degradation.

The core contribution of this work is showing that cross-model disagreement can function as an external, system-agnostic uncertainty proxy. In typical clinical ASR deployments, users receive a single best-hypothesis transcript with limited uncertainty information. The multi-model approach transforms this opaque output into a risk-stratified reading surface: positions where systems diverge encode observable ambiguity and can be surfaced for targeted human review. This workflow is reference-free in the sense of lacking a human-verified reference transcript; the consensus pseudo-reference is a deterministic aggregation used only to compute disagreement, not a correctness label.

The present findings relate to but differ from classical work on multi-system combination. Voting-based system combination approaches such as ROVER and confusion network methods have long leveraged agreement across recognizers to improve accuracy, typically evaluated against reference transcripts [12]. The present study instead uses agreement and disagreement primarily to surface and localize uncertainty without a reference, shifting the objective from minimizing average error to enabling triage-oriented verification. Similarly, ASR confidence estimation has often focused on model-internal signals and downstream calibration [10]. Cross-model disagreement is a black-box compatible uncertainty proxy: it requires no modification to any ASR system and remains available even when confidence scores are absent or incomparable.

**Relationship to the MEDLEY Paradigm**

These findings are consistent with the generalizability of the MEDLEY paradigm beyond diagnostic applications [14]. MEDLEY proposes that medical AI systems should orchestrate multiple models while preserving their diverse outputs rather than collapsing them into a consensus, treating disagreement as clinically informative [14]. Our ASR ensemble instantiates MEDLEY's four principles: **Diversity** (eight heterogeneous systems with low ICC = 0.131, supporting complementary failure modes); **Transparency** (systematic documentation of model provenance and performance); **Plurality** (risk-band framework preserving the full agreement distribution); and **Context** (accent-stratified analyses and differentiation of content from formatting disagreements).

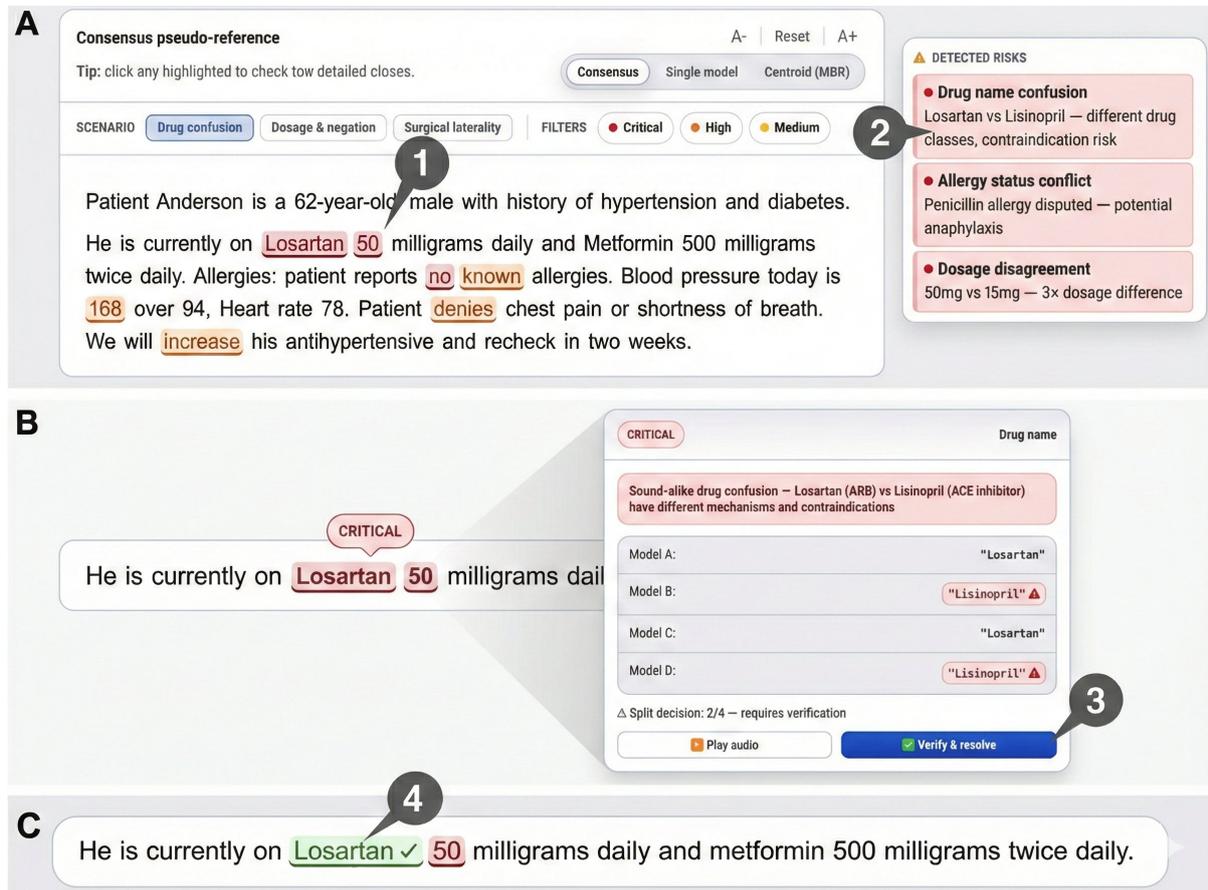

*Figure 5. Conceptual risk-aware disagreement review interface and orchestration workflow (future work)*. (A) Dashboard summarizing clip-level risk mass and a selectable review threshold alongside a risk-highlighted consensus transcript. (B) Token inspection view showing aligned alternatives from each ASR system with disagreement-category labels. (C) Review queue and audit panel supporting resolution, correction capture, and export for monitoring. Numbered callouts: (1) selected high-risk token, (2) linked risk-category item, (3) adjudication action with audit logging, (4) resolved indicator confirming closure.

MEDLEY's central thesis (i.e., that model imperfection can be reframed as a resource) finds support in our content-enrichment analysis: low-agreement regions preferentially capture lexical divergence in meaning-bearing categories. Both frameworks emphasize human-in-the-loop verification; a risk-aware disagreement review interface could surface flagged spans for manual review (Figure 5) rather than automatically accepting any single model's output, reinforcing the need for active clinician oversight when integrating AI-generated documentation into routine workflows [4].

**Future Work**

Future work should connect the risk map to external correctness through expert annotation of stratified subsets (e.g., a small, pre-registered sample of high-risk tokens and matched low-risk controls, adjudicated by clinicians), enabling estimation of how strongly each risk band predicts verified errors and whether enrichment holds across speaker groups. Such validation would establish whether content-category disagreement corresponds to actual transcription errors at rates sufficient to justify clinical deployment. A clinician-centered usability study is also needed to test workflow claims directly (e.g., review time, correction yield, and perceived workload) when disagreement highlighting is integrated into routine documentation review. Scaling the corpus to include real clinical encounters with turn-taking and ambient noise would enable assessment of ecological validity and recalibration of risk-band prevalence under deployment-like conditions.

To move from retrospective analysis toward deployable decision support, we outline a risk-aware disagreement review interface and an "orchestrator" layer that treats multi-ASR outputs as complementary evidence rather than collapsing them into a single transcript (Figure 5). The interface would visualize the consensus pseudo-reference with token-level risk bands derived from majority strength, emphasize clinically salient disagreement categories (e.g., numeric quantities, medication terms, and negation), and allow rapid adjudication by exposing aligned per-model alternatives with audit logging. The orchestrator would route unresolved high-risk spans to additional checks (e.g., targeted re-recognition or structured templates) and incorporate human corrections to recalibrate thresholds over time. This concept is presented as future work and was not implemented or evaluated in the current study.

## Conclusions

Cross-model disagreement provides a practical, reference-free signal for localizing transcription uncertainty in medical speech. The signal is sparse (concentrated in a small minority of tokens), systematic (driven by inter-model heterogeneity rather than purely stochastic variation), and enriched for content-category disagreements (i.e., lexical divergence in meaning-bearing word classes). These properties support its potential use as a triage mechanism for prioritizing human verification in future clinical transcription workflows, addressing the documented need for active physician oversight of AI-generated clinical documentation. We evaluate disagreement structure without human-verified transcripts, focusing on whether disagreement is sparse, localizable, and enriched for meaning-bearing lexical divergence. More broadly, these findings are consistent with the MEDLEY paradigm's central thesis (i.e., that cross-model disagreement can operationalized as a structured clinical resource), extending its applicability from diagnostic reasoning to medical documentation workflows.

## Acknowledgments

[To be added]

## Funding

[To be added]

## Competing Interests

The authors declare no competing interests.

## Author Contributions

[CRediT author statement to be added]

## Ethics Statement

This study involved secondary analysis of publicly available audio recordings obtained from open-access online sources. No interaction with human participants occurred, and no protected health information or personally identifiable information was collected or analyzed. The study was conducted in accordance with the principles of the Declaration of Helsinki. In accordance with Swedish regulations and institutional policy, this work did not constitute research involving human subjects and therefore did not require review or approval by a regional ethical review authority.

## Data Availability

The audio corpus was derived from publicly available YouTube videos. Analysis code and processed data files are available from the corresponding author upon reasonable request.

Multi-Model Disagreement for Medical Transcription